# A Historical Discussion of Angular Momentum and its Euler Equation

Amelia Carolina Sparavigna[1]

[1]Department of Applied Science and Technology, Politecnico di Torino, Italy

**Abstract:** We propose a discussion of angular momentum and its Euler equation, with the aim of giving a short outline of their history. This outline can be useful for teaching purposes too, to amend some problems that students can have in learning this important physical quantity.

## 1. Introduction

When we are teaching the laws concerning angular momentum to the students of a physics classroom, we are usually proposing them as the angular version of Newton's Laws, in particular when we are discussing the rotational motion. To illustrate it we need several concepts and physics quantities: angular velocity and acceleration, cross product of vectors, torques and moments of inertia. Therefore, the discussion of angular momentum is rather complex, and requiring a large number of examples in order to have the student confident with this important physical quantity.

Besides examples and exercises, a short history of this concept can help the students to highlight the links between celestial mechanics of planets and rigid body mechanics. In this paper then, we propose a discussion of angular momentum and its Euler's equation in the framework of a short outline of their history. Let us start from the concept of angular momentum and from an object well known from the ancient past, the spinning top.

## 2. Angular momentum and spinning tops

In physics, angular momentum is important because it is a conserved quantity. In fact, this physical quantity remains constant unless acted on by an external torque. Before given in its modern form by means of Noether's theorem [1], the conservation of the angular momentum was discussed in two manners, linked to the rotational inertia of bodies and to the motion of revolution of planets.

For what concerned rotational motion, an object was quite attractive in the past, the top and its endless spin. Isaac Newton wrote about it in the following passage of his Principia: "Projectiles persevere in their motions, so far as they are retarded by the resistance of the air, or impelled downwards by the force of the gravity. A top, whose parts, by their cohesion, are perpetually drawn aside from rectilinear motions, does not cease its rotation otherwise than it is retarded by the air". He also linked spinning tops and planets, telling that the "greater bodies of the planets and comets, meeting with less resistance in more free spaces, preserve their motions both progressive and circular for a much longer time" [2].

Of course, spinning tops had been mentioned in literature quite before Newton [3]. Plato talks of them in his Republic (360 BC) when he is discussing of rest and motion. He tells that when we see a man standing still but moving his hands and head, we can tell that this man is at the same time at rest and in motion. We observe the same in spinning tops. Tops, in a certain manner, stand still as a whole and, at the same time are in motion when, with the peg fixed in one point, they revolve. "The same is true of any other case of circular motion about the same spot" (of course we are not considering repose and movement in relation to the same parts of the objects). "We would say that there was a straight line and a circumference in them and that, in respect of the straight line, they are standing still since they do not incline to either side, but in respect of the circumference they move in a circle". When, during their revolution, tops "incline the perpendicular to right or left or forward or back, then they are in no wise at rest" [4]. A spinning top is therefore having an "equilibrium mobile" [4], which disappears when we have the precession, that is, when the axis of the top is moving.

Spinning top is appearing also in a Saint Basil's Hexaemeron (Homily 5) [3]. "Let the earth bring forth". It is this command (of God) "which, still at this day, is imposed on the earth, and in the course of each year displays all the strength of its power to produce herbs, seeds and trees. Like tops, which after the first impulse, continue their evolutions, turning upon themselves when once fixed in their centre; thus nature, receiving the impulse of this first command, follows without interruption the course of ages, until the consummation of all things" [5]. In fact, Basil's discussion seems a conservation law of nature.

During the Medieval period, we find spinning top mentioned by Jean Buridan (1295-1358) as a counterexample to Aristotelian "doctrine of antiperistasis" [6]. Then, the puzzling verticality of this rotating body appeared in the books of Italian Renaissance writers. Giovanni Battista Benedetti (1530-1590), for instance, is arguing that the reason for the spinning top to remain for some time erect over its tips is in the natural rectilinear tendency of it parts, which increasing with the speed of rotation, overwhelms the natural tendency downwards [6]. However, this tendency is never wholly obliterated by the tangential tendency, and then Benedetti argued that a body becomes lighter the more swiftly it is spinning [6].

After Newton, the angular momentum was considered in terms of the conservation of areal velocity, a result of the analysis of Kepler's Second Law of planetary motion [7]. During the motion of a planet, the line between the Sun and the planet sweeps out equal areas in equal intervals of time. Newton derived a proof of this law, using geometry and demonstrated that the attractive force of the Sun was the origin of all of Kepler's Laws.

As told in [7], many scientists and philosopher considered the conservation of the angular momentum mainly viewed like a conservation of areal velocity. In 1746, both Daniel Bernoulli (1700-1782) and Leonhard Euler (1707-1783) gave the proof of the conservation of the angular momentum (conservation of the momentum of rotational motion [8]) for a point-like mass sliding along a smooth tube in a horizontal plane [9]. As we will see in the following, Euler considered also the angular momentum of rigid bodies and proposed an equation for it in his works on mechanics [10,11].

We have to wait until 1803, to see a representation of the angular momentum similar to that we are using today with a cross product of vectors. In fact, that year, Louis Poinsot proposed representing the rotations as a line segment perpendicular to them. He also elaborated some discussions on the "conservation of moments" [7]. It was in 1858, in the William Rankine's Manual of Applied Mechanics, that we can find the angular momentum defined in the modern sense for the first time [7]. "A line whose length is proportional to the magnitude of the angular momentum, and whose direction is perpendicular to the plane of motion of the body and of the fixed point, and such that, when the motion of the body is viewed from the extremity of the line, the radius-vector of the body seems to have right-handed rotation". In 1872 edition of the same book, Rankine stated that the term "angular momentum" was introduced by R.B. Hayward [7]. Rankine was probably referring to Hayward's article "On a direct method of estimating velocities, accelerations, and all similar quantities with respect to axes moveable in any manner in space with applications" [7]. Hayward's article apparently was the first use of the term "angular momentum"; previously, angular momentum was typically referred to as "momentum of rotation" in English.

As previously told, Euler linked the idea of angular momentum to the rotation of the bodies. Before talking about Euler and his laws concerning rigid bodies, let us consider an important fact. We are used to define the angular momentum of a particle as a cross product of its position vector and its linear momentum. However, Euler, like Newton, did not use vectors in physics. They, of course, considered vectorial quantities but never the concept of a vector. The systematic study and use of vectors were a 19th and early 20th century phenomenon [12].

### 3. Straight motion and angular momentum

Sometimes, students are surprised that a particle moving on a straight line can have an "angular momentum". To help them, it is possible to use the polar coordinates to show that an "angular" variation exists and that, consequently, we have to observe an angular momentum also in this case. In fact, we can use a historical approach too, telling them that the position vector of the particle sweeps out equal areas in equal intervals of time, when the particle is moving on a straight line with a constant speed. Let us give a fixed origin O of position vectors in a plane containing the particle trajectory, as shown in the Figure 1; the position vector of the particle sweeps out equal area in equal time intervals.

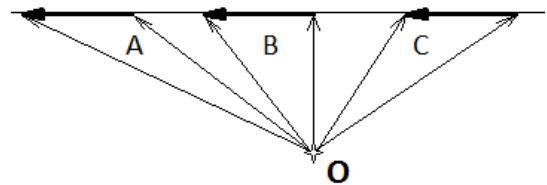

**Figure 1:** A particle is moving on a straight line with constant speed. The position vector of the particle sweeps out equal area in equal time intervals. The three triangle A, B and C, have the same base, given by speed times the interval of time, and the same altitude, therefore their areas are the same.

Another students' difficulty is the explicit calculation of the value of angular momentum, with respect to a given point in the space. Let us consider a Cartesian frame (x,y,z), and a mass m moving along the x-axis, with constant speed. Let us find the angular momentum of the particle with respect to a fixed point having coordinates (a,b,c) for instance. In a Poinsot or Rankine approach, the momentum is a line segment perpendicular to the plane of rotation, determined by position vector $\vec{r}$ and by x-axis, as shown in Figure 2. The position vector changes with time as the particle is moving, but, to evaluate the angular momentum, it is important its projection in the plane perpendicular to the axis of motion.

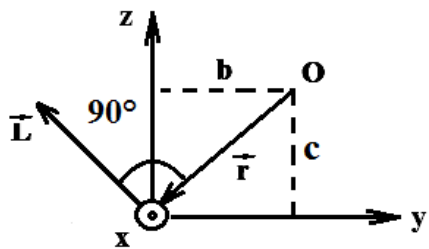

**Figure 2:** The angular momentum is perpendicular to the plane given by position and velocity vectors.

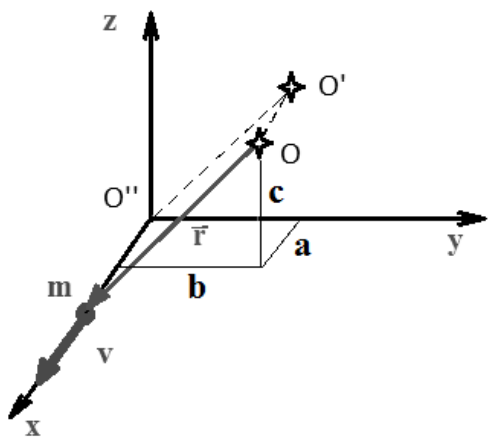

**Figure 3:** In the angular momentum, the position vector has its origin from a given point O. In the figure, this point has coordinates (a,b,c) in the space having the frame of reference with origin O". O' is the projection of O on the plane (y,z).

In the Figure 3 we can see O, its projection O' on the plane (y,z) and the origin O" of the frame of reference. The position vector, having its origin in O, gives the position of a particle, which, in our example, is moving on the x-axis with a constant speed. We have the angular momentum:

$$\vec{L}_O = \vec{r} \times m\vec{v} \quad (1)$$

However, we can write the position vector as the sum of three vectors:

$$\vec{r} = \overrightarrow{OP} = \overrightarrow{OO'} + \overrightarrow{O'O''} + \overrightarrow{O''P} \quad (2)$$

Vectors $\overrightarrow{OO'}$ and $\overrightarrow{O''P}$ are parallel to the velocity vector. Therefore, the only vector contributing to the cross product is $\overrightarrow{O'O''}$:

$$\vec{L}_O = \vec{r} \times m\vec{v} = \overrightarrow{O'O''} \times m\vec{v}$$

$$= m \begin{vmatrix} \hat{i} & \hat{j} & \hat{k} \\ 0 & -b & -c \\ v & 0 & 0 \end{vmatrix} = mv\left(-c\hat{j} + b\hat{k}\right) \quad (3)$$

This is the same result shown in the Figure 2.

### 4. Euler's equation

The laws concerning the motion of a rigid body are the Euler's Laws. Usually seen as the angular version of Newton's Laws, they were proposed by Euler in 1736, in his book entitled "Mechanica, sive motus scientia analytice exposita", about 50 years after Isaac Newton formulated his laws [11]. Euler was the first to discuss the general motion of a rigid body, showing that we can see it as motion of its "center of inertia" (center of mass), and rotation about an axis passing through it [8,10]. In fact, Euler started the development of his laws from the recognition that any infinitesimal motion of a body can be decomposed into a translation and a rotation [10].

One of the Euler's Laws tells that the angular momentum of a system of particles or of a rigid body changes because of the torques applied on it. The rate of change of the angular momentum depends only on the torques of external forces, because the sum of internal torques is zero. This law, governing the rate of change of angular momentum, is also known as "Balance Equation" [13]. In Italian Academy, we usually call it the "Second Cardinal Equation" [14]. The "First Cardinal Equation" is concerning the acceleration of the center of mass (see Appendix).

Using a Google search among books, we can see that the Euler's law is defined as "Second Cardinal Equation" in an Italian book of 1912 [15], and used for determining the equilibrium of bodies: "Per la seconda equazione cardinale dell'equilibrio, il momento del sistema delle forze motrici esterne rispetto ad ogni punto, come polo, deve essere nullo" [15]. It seems that this is the first book in which the Euler's law is called in this manner.

Of course, the Euler's law appear in older Italian book. For example, we have it in a book printed in 1828 [16], where it is discussed the problem of a balance. Being S the moment of inertia of the balance, about its rotation axis, the time derivative of the angular velocity $\omega$ is given by the torque divided by S. It is therefore clear the fact that this balance equation is connected to the ancient problem of equilibrium of torques.

In [17], we have at Page 133, the Euler's equation given in a Corollary; we are reproducing it in the following. "Coroll.2. But if the action of the force $F$ is continual, the rotatory motion will be accelerated, and we shall have

$$\frac{d\omega}{dt} = \frac{aF}{S}. \quad (4)$$

For, let $d\omega$ be the increment which the angular velocity receives, in the instant of time $dt$. The increment of velocity, which the element $dM$ receives, will be $r\,d\omega$; whence its accelerating force will be $r\,d\omega/dt$, and its moving force will be $r.d\omega.dM/dt$. Now, all these forces, acting in a contrary direction, must counterbalance the force $F$, and, therefore the sum of their momentums $= aF$. Hence $\frac{d\omega}{dt}\int r^2 dM = \frac{S\,d\omega}{dt} = aF$." [17].

### 5. Moment of inertia

The physical quantity S is the moment of inertia. In fact, in the Corollary we find the letter S to tell that it is a sum on the element of mass $dM$. We can find the moment of inertia in determining the motion of a boat [18], because Euler was involved in nautical engineering problems too (let us note that he called S the "momentum of inertia" [19]).

In the book of Venturoli [17] we can find also a discussion on the principal axes of inertia. In fact, Euler demonstrated that each body has three mutually orthogonal axes, the directions of which are defining the so-called principal axes of rotation. These axes are special because, when the body rotates about one of them, the angular momentum vector becomes parallel to the angular velocity vector. Moreover, a rigid body admits a stationary rotation about any one of the principal axes (a motion of a body, under which its angular velocity remains constant, is called a stationary rotation) [20].

Let us stress that, as told in [8], the theorem of the three principal axes had been already proposed and demonstrated, before Euler, by János András Segner (1704-1777). Actually, it was the study of Segner on the gyroscopic motion, which led Euler to the formulation of the laws of rigid bodies [21]. Moreover, besides introducing the moment of inertia, Segner also demonstrated that if the axis of rotation of a rigid body in rotary and translational motion, goes across its centre of mass, then the rotary and translational motions are independent from one another [21]. Besides these fundamental results, Segner invented a horizontal waterwheel, of which he discussed in some letter with Euler [22].

Euler was not to first to invent the moment or "momentum" of inertia then, but it seems he was the first to call it, in Latin, "momentum inertiae" [23]. Euler used this term in his book entitled "Theoria Motus Corporum Solidorum" [23]. "Ratio hujus denominationis ex similitudine motus progressive est desumta: quemadmodum enim in motu progressivo, si a vi secundum suam directionem sollicitante acceleretur, est incrementum celeritatis ut vis sollictitans divisa per massam seu inertiem; ita in motu gyratorio, quoniam loco ipsius vis sollicitantis ejus momentum considerari oportet, eam expressionem $\int r^2 dM$, quae loco inertiae in calculum ingreditur, momentum inertiae appellemus, ut incrementum celeritatis angularis simili modo proportionale fiat momento vis sollicitantis diviso per momentum inertiae." Euler also demonstrated that the moment of inertia of a rod, $L$ long, turning about a perpendicular axis passing through one of its ends is equal to $ML^2/3$, in a letter from Berlin, 20 May 1755 [24].

### 6. Conclusion

Let us conclude this discussion on the history of Euler equation, remembering that Euler and Segner were among those scientists that, working on theoretical and technological problems of the rotatory motion, paved the way for inventions and developments of the Industrial Revolution. Therefore, after discussing with students the Euler's Laws of motion, it would be interesting to illustrate them the use of the flywheels [25] in machines and steam engines of Industrial Revolution, to have a stronger link between the physics of fluids and thermodynamics to the Eulerian mechanics.

**Appendix**

The Second Cardinal Equation has a general form:

$$\frac{d\vec{L}_A}{dt} = \vec{\tau}_A^{(e)} - \vec{v}_A \times M\vec{v}_{CM} \quad (A1)$$

A is the origin of position vectors (pivot), not necessarily at rest in the inertial frame of reference. $\vec{v}_{CM}$ is the velocity of the centre of mass. M is the mass of the system of particles or of the rigid body. The torques relevant in this equation are only those of the external forces.

Usually, the pivot is chosen so that the term $\vec{v}_A \times M\vec{v}_{CM}$ is zero. This happens when $\vec{v}_A = 0$ or when $\vec{v}_{CM} = 0$. Moreover, it happened when pivot A and the centre of mass are coincident. The term is also null when the pivot and the centre of mass have the same velocity: $\vec{v}_A = \vec{v}_{CM}$.

This term exists: let us illustrate the role of $\vec{v}_A \times M\vec{v}_{CM}$ in a specific example. Let us have a disk rolling without slipping on a horizontal surface.

We could write the Cardinal Equation for three pivots for instance: a point A which is fixed on the horizontal surface, the point C of contact between disk and surface, and the center of mass CM of the disk.

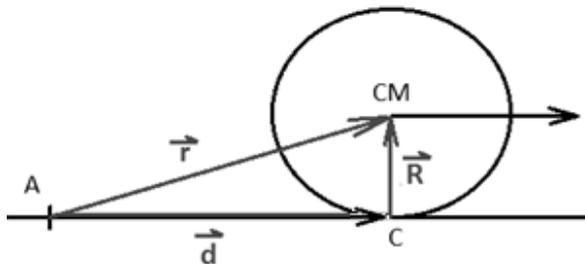

**Figure 4**: A disk is rolling without slipping on a horizontal surface. We can use three pivots for instance: A, C and CM.

Let us note that the contact point C is usually considered like a point "momentarily at rest", because there is no relative motion between disk and horizontal surface. In this manner, the disk is seen as "rotating" about an axis passing through C. However, C is changing its position with time at a speed equal to that of the center of mass. Then C has the same velocity of CM: $\vec{v}_C = \vec{v}_{CM}$. Therefore $\vec{v}_C \times M\vec{v}_{CM} = 0$.
Another Euler's Law, the Law that in Italian Academy we define as the First Cardinal Equation, is determining the motion of the center of mass. It is giving the acceleration of it in the following form:

$$M\vec{a}_{CM} = \vec{F}^{(e)} \quad (A2)$$

In this equation, only the external forces are relevant. Usually, to solve the problem of the motion of a rigid body, the First Equation (A1) is considered with the Second Equation (A1) given for pivot coincident with the center of mass.

**References**
1) Kleinert, H. (2009). Path Integrals in Quantum Mechanics, Statistics, Polymer Physics, and Financial Markets, World Scientific. ISBN: 978-981-4273-56-5
2) Newton, I. (1803). Axioms; or Laws of Motion, Law I. in The Mathematical Principles of Natural Philosophy. Andrew Motte translator. H. D. Symonds, London.
3) Oliver, V. (2002). History of the Top. Web page available at www.spintastics.com/ SSThisttop.html
4) Plato (1969). Plato in Twelve Volumes, Vols. 5 & 6. Paul Shorey translator. Cambridge, MA, Harvard University Press; London, William Heinemann Ltd. See Chapter 4, 436.
5) Saint Basil (1895). Hexaemeron, Jackson Blomfield translator; revised and edited by Kevin Knight. In From Nicene and Post-Nicene Fathers, Second Series, Vol. 8, Philip Schaff and Henry Wace Editors. Buffalo, NY, Christian Literature Publishing Co.
6) Gabbey, A. (1990). The Case of Mechanics: One Revolution or Many? In Reappraisals of the Scientific Revolution, David C. Lindberg and Robert S. Westman Editors, Cambridge University Press, pp. 493-528. ISBN: 9780521348041
7) Vv. Aa. (2015). Angular Momentum, in Wikipedia and references therein.
8) Andrés, G. (1790). Dell'Origine, Progressi e Stato Attuale d'Ogni Letteratura: Della Meccanica, Volume 4, Stamperia Reale. Google e-book.
9) Caparrini, S.; Fraser, C. (2013). Mechanics in the Eighteenth Century. In The Oxford Handbook of the History of Physics, Jed Z. Buchwald and Robert Fox Editors, OUP, Oxford, pp. 358-405. ISBN: 9780199696253
10) Borrelli, A. (2011). Angular Momentum Between Physics and Mathematics. In Mathematics Meets Physics, Karl-Heinz Schlote and Martina Schneider Editors, Verlag Harri Deutsch, Frankfurt a.M., pp. 395-440. ISBN: 978-3-8171-1844-1
11) McGill, D.J.; King, W.W. (1995). Engineering Mechanics, An Introduction to Dynamics, PWS Publishing Company. ISBN: 0-534-93399-8
12) Crowe, M.J. (1967). A History of Vector Analysis: The Evolution of the Idea of a Vectorial System, Courier Corporation. ISBN: 9780486679105
13) Hutter, K. (2014). Continuum Mechanics in Environmental Sciences and Geophysics, Springer. ISBN: 978-3-7091-2600-4, DOI: 10.1007/978-3-7091-2600-4
14) Romano, A. (2012). Classical Mechanics with Mathematica®, Springer Science & Business Media, Birkhäuser Basel. ISBN: 978-0-8176-8352-8, DOI: 10.1007/978-0-8176-8352-8
15) Maggi, G.A. (1912). Dinamica Fisica: Lezioni sulle Leggi Generali del Movimento dei Corpi Naturali, E. Spoerri, Pisa.
16) Masetti, G.B. (1827). Note ed Aggiunte agli Elementi di Meccanica e d'Idraulica del Professore Giuseppe Venturoli, Volume 2, Tipografia Cardinali e Frulli, Bologna. Google e-book.
17) Venturoli, G. (1822). Elements of the Theory of Mechanics, Translated from the Italian by D. Cresswell, J. Nicholson & Son, Cambridge. Google e-book.
18) Vv. Aa. (1784). Memorie di Matematica e Fisica della Società Italiana, Volume 2, 1784, Dionigi Ramanzini, Bologna. See Cap.3, p. 467. Google e-book.
19) Wildbore, C. (1790). On Spherical Motion, communicated by Earl Stanhope, on 24 June 1790 to the Philosophical Transactions of the Royal Society of London, Volume 80, W. Bowyer and J. Nichols for Lockyer Davis, printer to the Royal Society. Google e-book. http://dx.doi.org/10.1098/rstl.1790.0028
20) Arnol'd, V.I. (1989). Mathematical Methods of Classical Mechanics, Springer Science & Business Media. ISBN: 978-1-4757-2063-1, DOI: 10.1007/978-1-4757-2063-1
21) Youschkevitch, A.P.; Grigorian, A.T. (2008). Segner, János - András (Johann Andreas von), in Complete Dictionary of Scientific Biography, Charles Scribner's Sons. ISBN: 0-684-31559-9
22) Reynolds, T.S. (2002). Stronger Than a Hundred Men: A History of the Vertical Water Wheel, JHU Press. ISBN: 9780801872488
23) Walton, W. (1842). A Collection of Problems in Illustration of the Principles of Theoretical Mechanics, Cambridge, W.P. Grant.
24) Vv. Aa. (1779). Vita Admodvm Reverendi Ac Magnifici Viri Iosephi Stepling, Sumptibus Caes. Reg. Scholae Normalis. Typis, per Ioannem Adamum Hagen.
25) Vv. Aa. (2015). Flywheel, in Wikipedia and references therein.